\documentclass[final,1p,times]{elsarticle} 
\usepackage{graphicx} 
\usepackage{amssymb} 
\usepackage{amsthm} 
\usepackage{lineno} 

\journal{Nuclear Physics A} 
\begin{document} 

\begin{frontmatter} 


\title{Gamma-Jet Tomography of Quark-Gluon Plasma in High-Energy Nuclear Collisions}

\author{Hanzhong Zhang$^{a,b}$ J. F. Owens$^{c}$ Enke Wang$^{a,b}$ and Xin-Nian Wang$^{d}$}

\address[a]{Institute of Particle Physics, Huazhong Normal University, Wuhan 430079, China}
\address[b]{Key Laboratory of Quark and Lepton Physics (Huazhong Normal University),
Ministry of Education, China}
\address[c]{Physics Department, Florida State University, Tallahassee, Florida 32306-4350, USA}
\address[d]{Nuclear Science Division, Lawrence Berkeley Laboratory, Berkeley, California 94720, USA}

\begin{abstract} 
Within the next-to-leading order (NLO) perturbative QCD (pQCD)
parton model, suppression of away-side hadron spectra associated
with a high $p_{T}$ photon due to parton energy loss is shown to
provide a complete tomographic picture of the dense matter formed in
high-energy heavy-ion collisions at RHIC. Dictated by the shape of
the $\gamma$-triggered jet spectrum in NLO pQCD, hadron spectra at
large $z_T=p_T^{h}/p_T^{\gamma} \stackrel{>}{\sim} 1$ are more
susceptible to parton energy loss and therefore are dominated by
surface emission of $\gamma$-triggered jets, whereas small $z_{T}$
hadrons mainly come from fragmentation of jets with reduced energy
from volume emission. These lead to different centrality dependence
of the hadron suppression in different regions of $z_{T}$.
\end{abstract} 

\end{frontmatter} 




Jet quenching \cite{wg90} has become a powerful tool for the study
of the quark-gluon plasma \cite{review} in high-energy nuclear
collisions. Jets are produced in the early stage of heavy-ion
collisions through hard parton scattering. When they pass through
the dense matter, they will interact with the medium and lose a
significant amount of their energy via gluon radiation induced by
multiple scattering.

In our previous studies based on a NLO pQCD parton model
\cite{owens87,zoww07}, we checked the tomographic pictures of single
jets and di-jets by a simultaneous fit to single hadron and dihadron
data. Single hadrons are dominated by the jet emissions close and
perpendicular to the surface of the system, while dihadrons are
emitted both close and tangential to the surface of the system
although there are contributions from punch-through jets from the
central region. However, the dominance of surface and tangential
emission makes it difficult to extract the space-time profile of the
dense medium from single and dihadron spectra.

Here we focus on the study of photon-triggered away-side hadron
spectra coming from $\gamma$-jet events in central nucleus-nucleus
collisions \cite{zoww09}. By selecting $\gamma$-hadron pairs with
different values of $z_{T}=p_{T}^h/p_{T}^{\gamma}$ which could be
larger than 1 due to radiative correction in NLO pQCD, one can
effectively control hadron emission from different regions of the
dense medium and therefore extract the corresponding jet quenching
parameters. For the study of photon-hadron correlation in this
paper, we focus mainly on photon production with isolation cuts
\cite{zoww09,owens90}. Therefore, we can neglect those photons that
are produced via induced bremsstrahlung \cite{vitevzhang},
jet-photon conversion \cite{Fries} and thermal production
\cite{Srivastava,TGJM} in high-energy heavy-ion collisions.

\begin{figure}
  \includegraphics[height=.29\textheight]{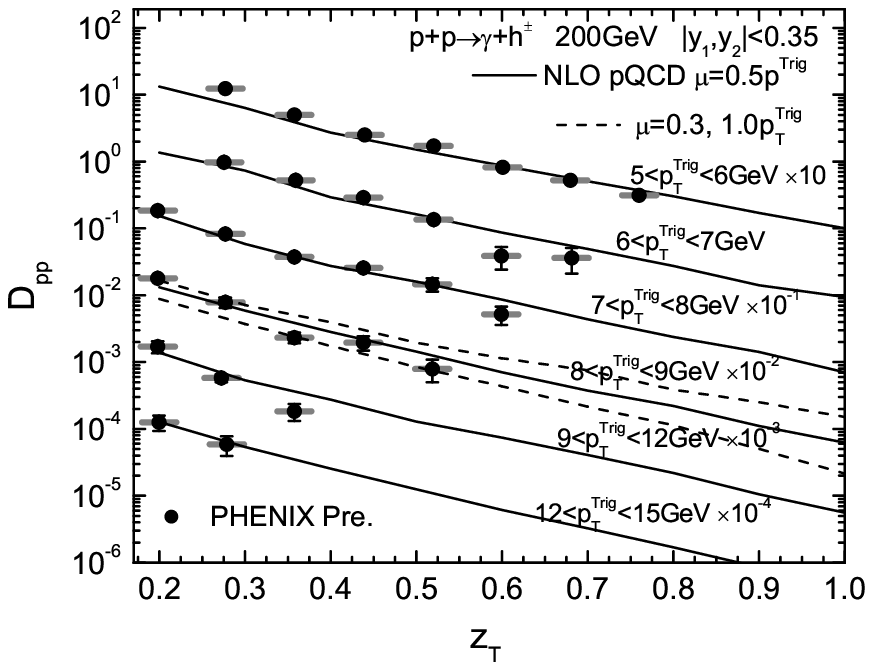}
  \includegraphics[height=.29\textheight]{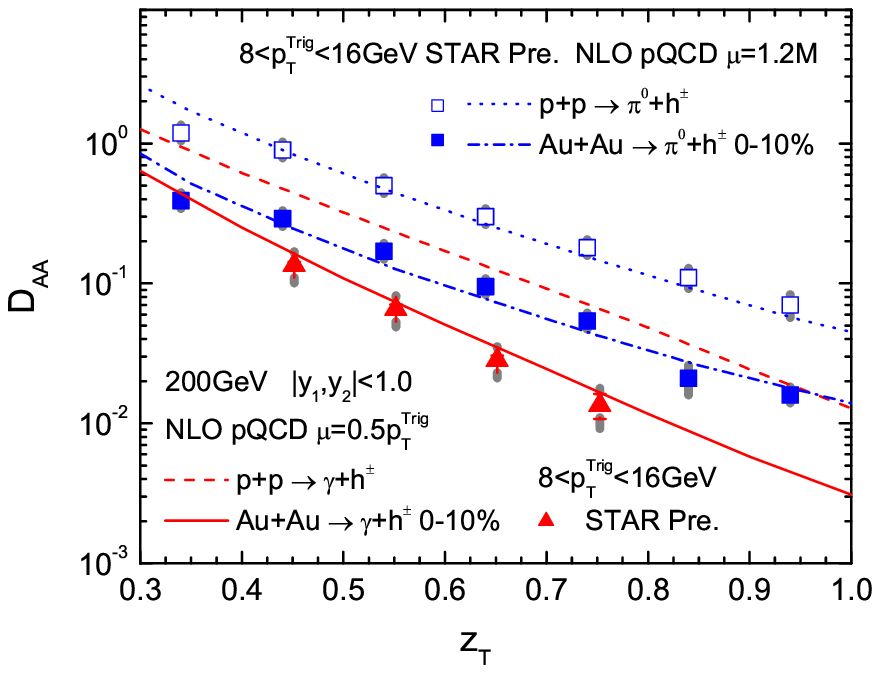}
  \caption{$\gamma$-triggered and hadron-triggered FF's in p+p and central Au+Au
collisions at the RHIC energy. The preliminary data are from
\cite{star-gam-hadr,frantz}. Systematic errors of experimental data
are shown as shaded bars when available.} \label{fig:Dpp_Daa}
\end{figure}

Within the same energy loss formalism as in our previous studies on
single and dihadron \cite{zoww07}, we calculate the production of
the photon-triggered hadron spectrum in central Au+Au collisions at
$\sqrt s$ = 200 GeV. Shown in Figure \ref{fig:Dpp_Daa} are our
numerical results for $D_{pp}(z_T)$ or $D_{AA}(z_T)$ \cite{zoww09}
compared to data. The left plot is for gamma-hadron spectra in p+p
collisions. They fit the PHENIX preliminary data very well for
different values of $p_T^{trig}$. We also show the uncertainties
(dashed curves) due to the choice of the factorization scale, which
mainly comes from the scale dependence of the FF's. In the right
plot we show the gamma-hadron spectra in central Au+Au collisions
(solid curve) as compared to p+p collisions (red dashed curve). The
NLO pQCD prediction for the suppression of gamma-hadron spectra
agrees well with the STAR preliminary data. Such an agreement is
extremely nontrivial given the completely different emission
geometry as compared to single and dihadron productions as we will
see below, and it reinforces the success of the parton energy loss
picture for the observed jet quenching phenomena. Also shown in the
right plot are the calculated hadron-triggered FF's, the dotted
curve for p+p collisions and the dot-dashed curve for central Au+Au
collisions, as compared to the experimental data. Hadron-triggered
FF's are larger than photon-triggered FF's for both p+p and A+A
collisions mainly because the fraction of hadron-triggered gluon
jets is larger than the fraction of photon-triggered gluon jets at
same $p_T^{trig}$, and the hadron yield of gluon jets is larger than
that of quarks.

\begin{figure}
  \includegraphics[height=.28\textheight]{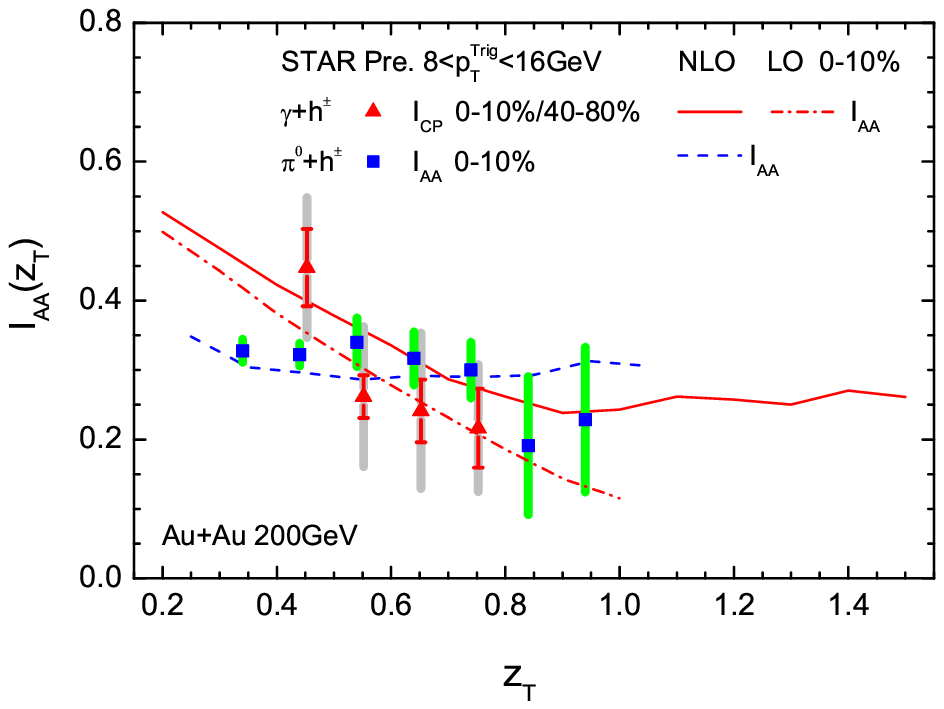}
  \includegraphics[height=.27\textheight]{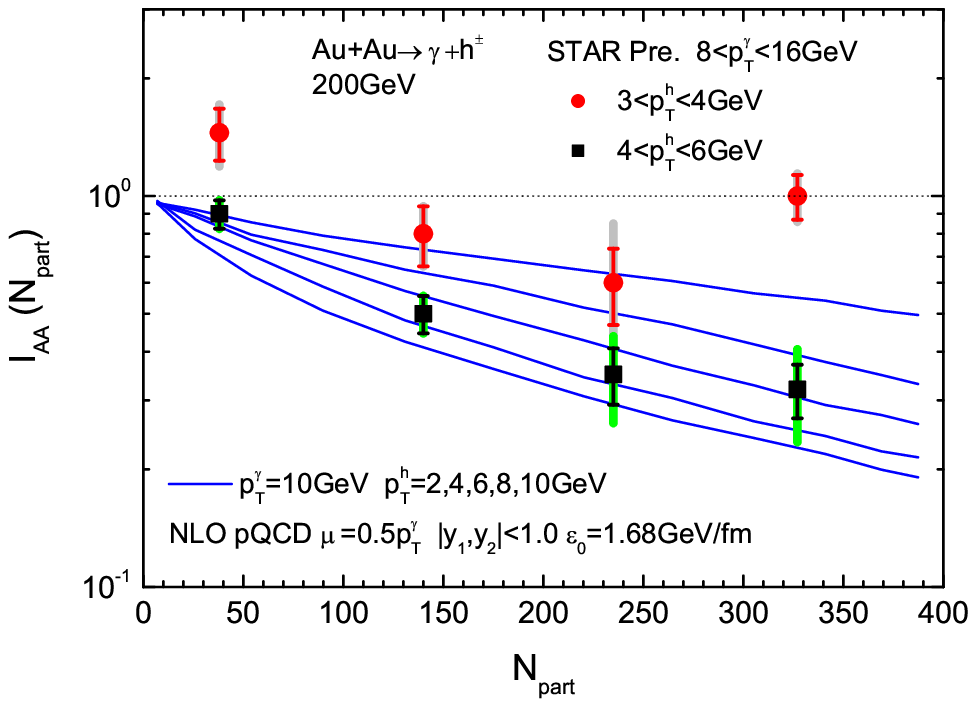}
  \caption{The nuclear
modification factor $I_{AA}$ for $\gamma$-triggered fragmentation
function in central $Au+Au$ collisions at the RHIC energy. The data
are from \cite{star-gam-hadr}. In the left plot for $I_{AA}(z_T)$
the triggered photon is chosen with $p_T^{\gamma}=8-16$ GeV. In the
right plot for $I_{AA}(N_{part})$ the triggered photon is chosen
with $p_T^{\gamma}=10$ GeV and the associated hadron with
$p_T^{\gamma}=2, 4, 6, 8, 10$ GeV, respectively (from top to
bottom).} \label{fig:Iaa}
\end{figure}

The nuclear modification factor $I_{AA}$ for the photon-triggered
hadrons $I_{AA}={D_{AA}}/{D_{pp}}$ is defined to characterize the
effect of jet quenching. Shown in the left plot of
Figure~\ref{fig:Iaa} are the calculated nuclear modification factors
both in LO (dot-dashed) and NLO (solid) calculations of
gamma-triggered FF's. In the LO pQCD calculation, transverse
momentum of the associated jet is balanced exactly by the direct
photon in tree $2 \rightarrow 2$ processes. This limits
$z_{T}=p_{T}^{h}/p_{T}^{\gamma}\leq 1$. In the NLO, however,
radiative correction permits $z_T>1$ for gamma-hadron productions.
The two effects give rise to NLO $I_{AA}$ very different from LO
results. Therefore, to exactly probe the dense matter from
gamma-hadron correlations, one must use NLO pQCD calculations. .

\begin{figure}
  \includegraphics[height=.285\textheight]{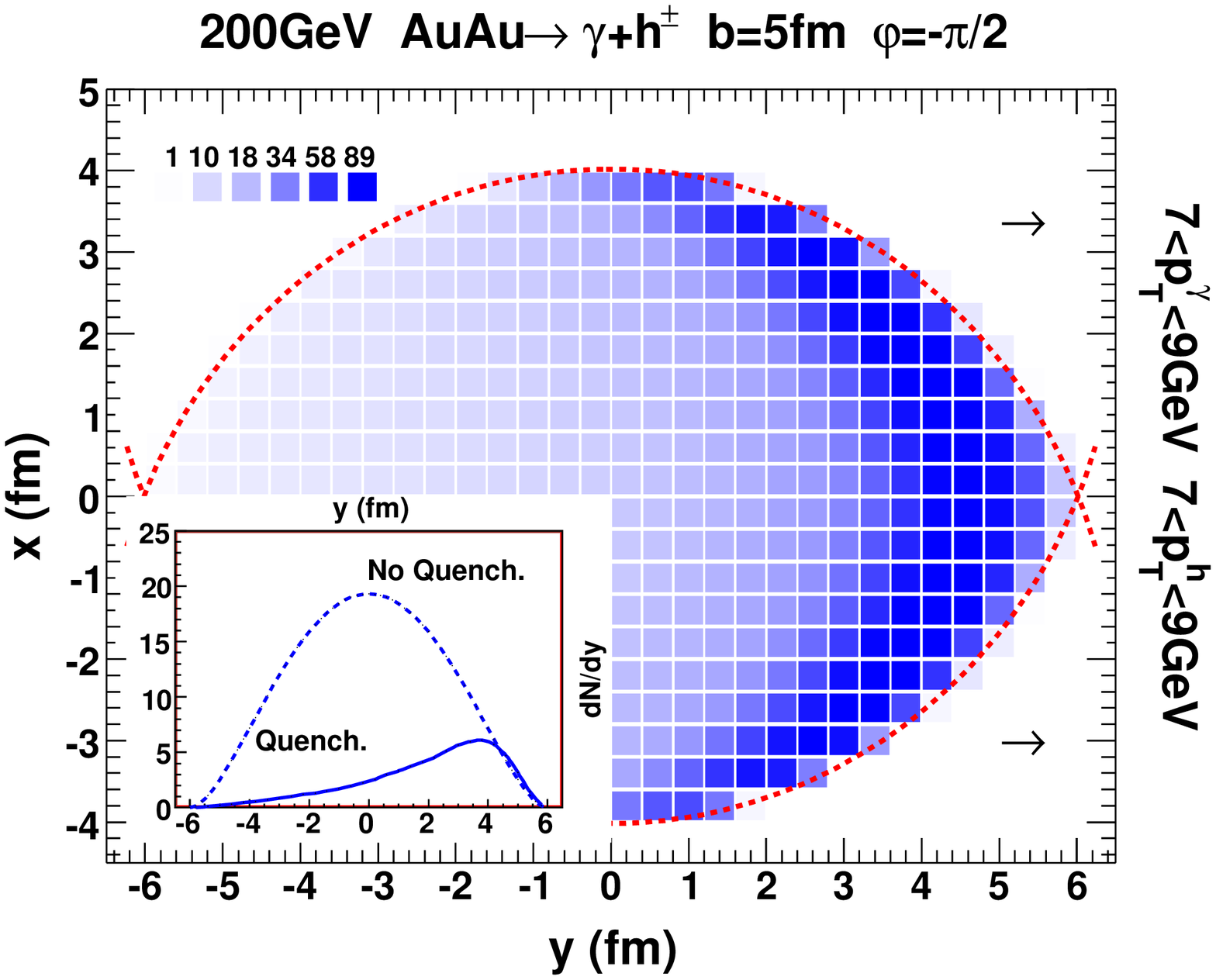}
  \includegraphics[height=.285\textheight]{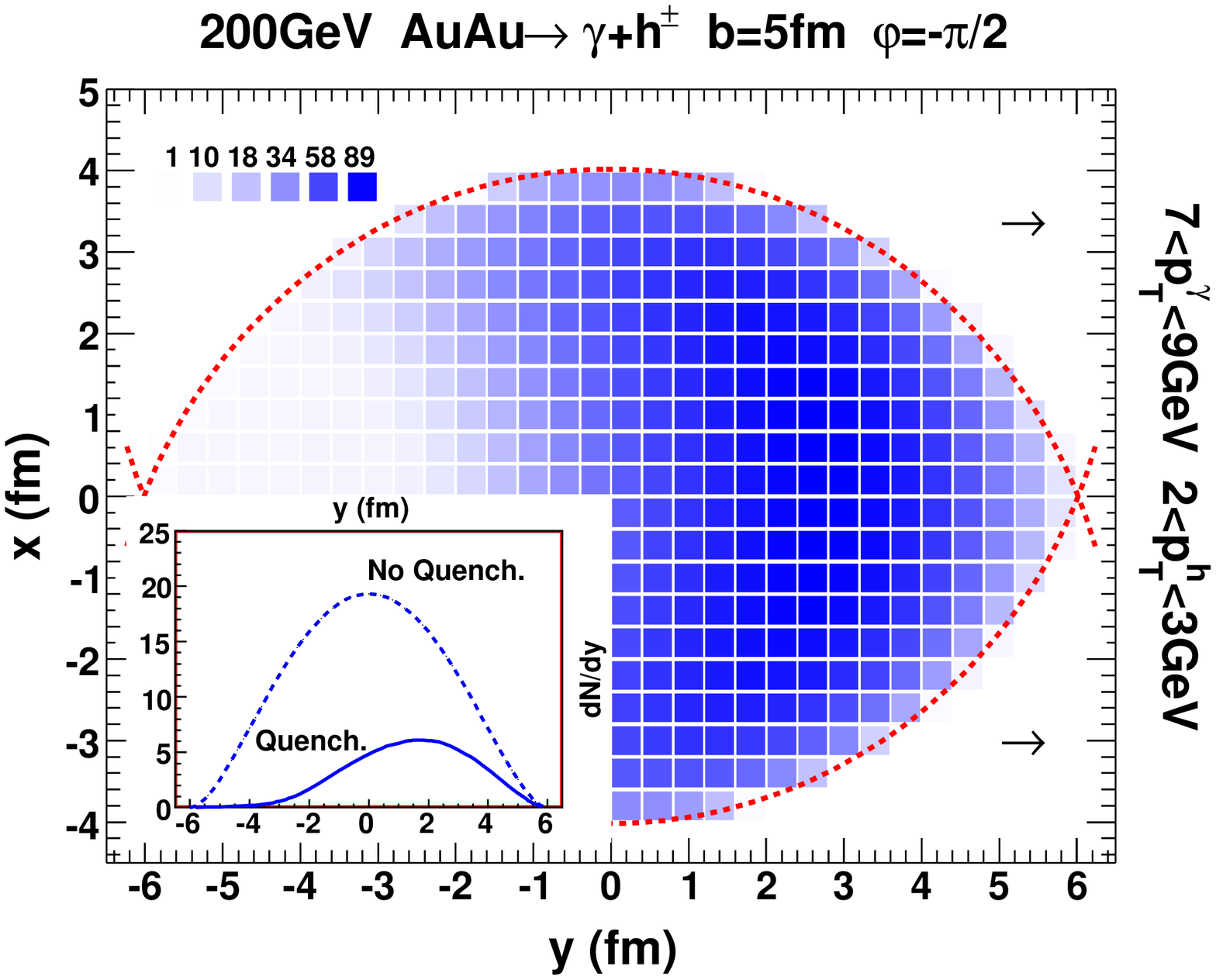}
  \caption{Transverse distributions
of the initial $\gamma$-jet production vertices that contribute to
the final observed $\gamma$-hadron pairs along a given direction
(arrows) with $z_T\approx 0.9$ (left plot) and $z_T\approx 0.3$
(right plot).} \label{fig:contourplot}
\end{figure}

Also shown in the left plot of Figure~\ref{fig:Iaa} is the dihadron
suppression factor, dashed curve. Compared with dihadron $I_{AA}$,
gamma-hadron $I_{AA}$ has a more stronger dependence on $z_T$. One
can imagine that the gamma-triggered jets contributing to
large-$z_T$ gamma-hadron are more susceptible to energy loss. Even a
small amount energy loss can greatly suppress the large-$z_T$
gamma-hadron yield. So large-$z_T$ gamma-hadrons are dominated by
those gamma-triggered jets originating near and escaping through the
surface almost without energy loss. Similar to single hadron
suppression factor, large-$z_T$ $I_{AA}$ is mainly determined by the
thickness of the corona of the surface emission. The picture of the
surface emission is demonstrated by the left plot of
Figure~\ref{fig:contourplot}. The plot is the spatial transverse
distribution of the initial gamma-jet production vertexes that
contribute to the final gamma-hadron pairs with given values of
$z_T$. The associated jets are considered along the right direction,
and the opposite direction is for the triggered photons. The color
strength represents the gamma-hadron yield from the fragmentation of
the gamma-triggered jets after parton energy loss. The inserted
panels are projections of the contour plots onto y-axes, solid curve
without energy loss, dashed curve with energy loss. As for small
$z_T$ region, it has contributions from energetic jets originated
from inside the medium that have lost a finite amount of energy
before fragmenting into hadrons. That's why these gamma-correlated
hadrons come from volume emission, as shown in the right plot of
Figure~\ref{fig:contourplot}. For the intermediate-$z_T$ region,
gamma-hadrons are determined by the competition of the two emission
mechanisms.

The above picture of volume and surface emission for the
$\gamma$-triggered fragmentation function in heavy-ion collisions
will lead to different centrality dependence of the nuclear
modification factor $I_{AA}(z_{T})$ in different regions of $z_{T}$.
Shown in the right plot of Figure~\ref{fig:Iaa} are nuclear
modification factors for $\gamma$-triggered hadron spectra as
functions of the participant number in $Au+Au$ collisions at the
RHIC energy for different values of $z_{T}$ as compared to the STAR
preliminary data. For small values of $z_{T}<1$, the
$\gamma$-triggered hadron yield is dominated by volume emission and
therefore the centrality dependence of the nuclear modification
factor is stronger than that in the region $z_{T}\geq 1$ where
surface emission is the dominant production mechanism.

In summary, high $p_T$ photon-hadron correlations are studied within
the NLO pQCD parton model with modified parton fragmentation
functions due to jet quenching in high energy $A+A$ collisions. We
demonstrated that the volume (surface) emission dominates the
$\gamma$-triggered hadrons spectra at small $z_{T}<1$ (large
$z_{T}\geq 1$) due to the underlying jet spectra in the NLO pQCD.
Therefore, one will be able to extract jet quenching parameters from
different regions of the dense medium by measuring the nuclear
modification factor of the $\gamma$-triggered fragmentation function
in the whole kinetic region, including $z_{T}\geq 1$, achieving a
true tomographic study of the dense medium.

\section*{Acknowledgements} This work was supported by DOE under
contracts DE-AC02-05CH11231 and DEFG02- 97IR40122, by NSFC of China
under Projects No. 10825523 and No. 10875052 and No. 10635020, by
MOE of China under Projects No. IRT0624; by MOST of China under
Project No. 2008CB317106; and by MOE and SAFEA of China under
Project No. PITDU-B08033.


\begin{thebibliography}{00} 

\bibitem{wg90}
X.~N.~Wang and M.~Gyulassy,
  Phys.\ Rev.\ Lett.\  {\bf 68}, 1480 (1992).

\bibitem{review}
  M.~Gyulassy, I.~Vitev, X.~N.~Wang and B.~W.~Zhang,
  arXiv:nucl-th/0302077;
  A.~Kovner and U.~A.~Wiedemann,
  arXiv:hep-ph/0304151,
   in Quark Gluon Plasma 3, eds. R. C. Hwa and X.N. Wang, World Scientific, Singapore, 2003.

\bibitem{owens87}
J.~F.~Owens, 
Rev.\ Mod.\ Phys. 59, (1987)465.

\bibitem{zoww07}
  H.~Z.~Zhang, J.~F.~Owens, E.~Wang and X.~N.~Wang,
  Phys.\ Rev.\ Lett.\  {\bf 98}, 212301 (2007).

\bibitem{zoww09}
  H.~Z.~Zhang, J.~F.~Owens, E.~Wang and X.~N.~Wang,
  Phys.\ Rev.\ Lett.\  {\bf 103}, 032302 (2009).

\bibitem{owens90}
H.~Baer, J.~Ohnemus, and
J.~F.~Owens, 
Phys.\ Rev.\ D {\bf 42}, 61 (1990).

\bibitem{vitevzhang}
I.~Vitev and B.~W.~Zhang,
  Phys.\ Lett.\  B {\bf 669}, 337 (2008).

\bibitem{Fries}
R.~J.~Fries, B.~M¨¹ller and D.~K.~Srivastava,
Phys.\ Rev.\ Lett.\  {\bf 90}, 132301 (2003).

\bibitem{Srivastava}
  D.~K.~Srivastava,
  J.\ Phys.\ G {\bf 35}, 104026 (2008).

\bibitem{TGJM}
S.~Turbide, C.~Gale, S.~Jeon and G.~Moore ,
Phys.\ Rev.\ C.\ {\bf 72}, 014906 (2005).

\bibitem{star-gam-hadr}
J.~Adams {\it et al.} ,
  Phys.\ Rev.\ Lett.\  {\bf 97}, 162301 (2006);
A.~M.~Hamed,
  J.\ Phys.\ G {\bf 35}, 104120 (2008);
  arXiv:0809.1462.

\bibitem{frantz}
  J.~Frantz  [PHENIX Collaboration],
  arXiv:0901.1393.




\end{thebibliography}
\end{document}